\documentstyle[11pt,epsf,fleqn]{elsart}
\renewcommand{\thefootnote}{\fnsymbol{footnote}}
\newcommand{\be}[1]{\begin{equation} \label{(#1)}}
\newcommand{\ee}{\end{equation}}
\newcommand{\ba}[1]{\begin{eqnarray} \label{(#1)}}
\newcommand{\ea}{\end{eqnarray}}
\newcommand{\nn}{\nonumber}
\newcommand{\rf}[1]{(\ref{(#1)})}
\def \znbb {$0\nu\beta\beta$}

\begin{document}
\begin{frontmatter}
\title{{\bf A superformula for neutrinoless double beta decay II:
The short range part}}
\author{H. P\"as $^1$},
\author{M. Hirsch $^2$},
\author{H.V. Klapdor--Kleingrothaus $^1$},
\author{S. G. Kovalenko $^3$}
\address{$^1$
Max--Planck--Institut f\"ur Kernphysik,\\
P.O.Box 10 39 80, D--69029 Heidelberg, Germany}
\address{$^2$ Inst. de Fisica Corpuscular - C.S.I.C.
- Dept. de Fisica Teorica, Univ. de Valencia, 46100 Burjassot, Valencia,
Spain}
\address{$^3$ Departamento de Fisica, Universidad Tecnica Federico
Santa Maria, Casilla 110-V, Valparaiso, Chile}
\begin{abstract}
A general Lorentz--invariant parameterization for the short-range
part of the $0\nu\beta\beta$ decay rate is derived. Combined with
the long range part already published this general parameterization
in terms of
effective $B-L$ violating couplings allows
one
to extract the
$0\nu\beta\beta$ limits on arbitrary lepton number violating theories.
 \end{abstract}
\begin{keyword}
Double beta decay; Neutrino; QRPA
\end{keyword}
\end{frontmatter}

\section{Introduction}
The search for neutrinoless
double beta decay has been proven to be one of the most powerful
tools to constrain $B-L$ violating physics beyond the standard model
\cite{Kla97}. In recent years, besides the most restrictive limit on
the effective neutrino Majorana mass \cite{Kla97,HM97,smirn},
stringent
constraints on several theories beyond the Standard Model such as
$R$--parity violating \cite{babu95,Hir96,Paes98} as well as conserving
\cite{Hir97} SUSY, leptoquarks \cite{Hir96b} and
left--right symmetric models \cite{Hir96c}
have been derived (for a review see \cite{Kla97}).
While the neutrino mass limit is based on the well-known mechanism
exchanging a massive Majorana neutrino between two standard model $V-A$
vertices, the effective vertices appearing in the new contributions
involve non--standard currents such as scalar, pseudoscalar and tensor
currents. Moreover, besides contributions with a light
neutrino being exchanged between two separated vertices, the so-called
long-range part,
additional contributions
from the short range part are expected, e.g. in SUSY without R-parity
\cite{Hir95,Hir95b}.

Thus we felt motivated to consider the neutrinoless double beta decay
rate in a general framework, parameterizing the new physics contributions
in terms of all effective low-energy currents allowed by Lorentz-invariance
(see Fig. 1).
Such an ansatz allows one to separate the nuclear physics part of double
beta decay from the underlying particle physics model, and to
derive limits
on arbitrary lepton number violating theories.
The first step of this work, treating the long--range part (contributions a-c in Fig. 1), has been published
in \cite{supf}. In the following the remaining part,
treating the short range part (contribution d), is presented.

Although the general decay rate is independent of the underlying
nuclear physics model, to extract quantitative limits, values of nuclear
matrix elements are needed, which have been calculated in \cite{Hir95b}
in the (pn)QRPA aproach.
Using these, for the first time limits for all potential new
physics contributions are derived here.

\section{General Parametrization}

In the short range part the effective interaction can be considered as
point-like, thus the decay rate results from first order perturbation
theory,
\ba{1}
R_{0\nu\beta\beta}&=&
i \int d^4 x
\langle (Z+2,A), 2e^-|{\cal L}(x)|(Z,A),-\rangle.
\ea
The most general Lorentz invariant Lagrangian has the following form
\ba{2}
&&
{\cal L}= \frac{G^2_F}{2} m_p^{-1} \{
\epsilon_1 J J j
+\epsilon_2 J^{\mu\nu} J_{\mu\nu} j
+\epsilon_3 J^{\mu}J_{\mu} j
+\epsilon_4 J^{\mu}J_{\mu\nu} j^{\nu}
+\epsilon_5 J^{\mu} J j_{\mu}     \nn \\ &&
+\epsilon_6 J^{\mu} J^{\nu} j_{\mu\nu}
+\epsilon_7 J J^{\mu\nu} j_{\mu\nu}
+\epsilon_8 J_{\mu \alpha} J^{\nu \alpha} j^{\mu}_{\nu} \},
\ea
with the proton mass $m_p$ and the hadronic currents of defined chirality
$J=\overline{u}(1\pm \gamma_5)d $,
$J^{\mu}=\overline{u} \gamma^{\mu} (1\pm \gamma_5)d$,
$J^{\mu\nu}=\overline{u} \frac{i}{2}[\gamma^{\mu},\gamma^{\nu}]
(1\pm \gamma_5)d$
and the leptonic curents
$j=\overline{e}(1\pm \gamma_5) e^C$,
$j^{\mu}=\overline{e}\gamma^{\mu} (1\pm \gamma_5) e^C$ ,
$j^{\mu\nu}=\overline{e}\frac{i}{2}[\gamma^{\mu},\gamma^{\nu}]
(1\pm \gamma_5) e^C$.
In some of the cases the decay rate for the effective coupling
$\epsilon_{\alpha}$ may
depend also on the chirality of the currents involved. In these cases we
define $\epsilon_{\alpha}=\epsilon^{xyz}_{\alpha}$, where $xyz=L/R,L/R,L/R$
defines the chirality of the hadronic and leptonic currents in the order of
appearance in eq. (2). In the cases where it is not necessary to distinguish
the different chiralities we suppress this additional index.

In renormalizable theories
no fundamental tensors exist. Thus the tensor currents have
to result either from
Fierz rearrangements or by integrating out heavy particles,
when deriving the effective Lagrangian
from the fundamental theory and decomposing
the expressions obtained in terms of the Lorentz invariant bilinears used
above, e.g. from
$\bar{u} \gamma_\mu \gamma_\nu (1+\gamma_5)d=
g_{\mu\nu} J  - i J_{\mu\nu}$.
Moreover, the contribution of the leptonic tensor current vanishes in
s-wave approximation. Since the tensor currents in the
rearrangements and decompositions mentioned above
are always created together with dominant contributions,
the corresponding terms proportional to
$\epsilon_6, \epsilon_7, \epsilon_8$ can be neglected.
The remaining terms are the ones proportional to
$\epsilon_1,\epsilon_2,\epsilon_3,\epsilon_4,\epsilon_5$.

Applying the standard nuclear theory methods based on
the non-relativistic impulse approximation we 
derive the general \znbb-decay half-life formula in s--wave approximation
%
\ba{t12}
&&[T_{1/2}^{0\nu\beta\beta}]^{-1}= \\ \nn
&&G_{1} |\sum_{i=1}^3\epsilon_i {\cal M}_i|^2 +
G_{2} |\sum_{i=4}^5\epsilon_i {\cal M}_i|^2 +
G_{3} {\rm Re}[(\sum_{i=1}^3\epsilon_i {\cal M}_i)
(\sum_{i=4}^5\epsilon_i {\cal M}_i)^*].
\ea
Here the phase space factors are
%
\ba{phase}
G_{1} = G_{01}, \ \ \
G_{2} = \frac{(m_e R)^2}{8} G_{09}, \ \ \
G_{3} = \left(\frac{m_e R}{4}\right) G_{06},
\ea
with $G_{0k}$ given in \cite{Doi85} and shown in Table 1.

The nuclear matrix elements in eq. \rf{t12} are
\ba{nucl1}\nn
&&{\cal M}_1 = -\alpha_1 {\cal M}_{F,N},\ \ \
{\cal M}_2 = -\alpha_2 {\cal M}_{GT,N},\\
&&{\cal M}_3 =
\frac{m_A^2}{m_P m_e}\{{\cal M}_{GT,N} \mp \alpha_3 {\cal M}_{F,N} \},\\
\nn
&&{\cal M}_4 = \pm\alpha_4 {\cal M}_{GT,N},\ \ \
{\cal M}_5 = \mp\alpha_5 {\cal M}_{F,N}.
\ea
In the last three cases contractions of
hadronic currents with different chiralities lead to different results.
The negative sign in ${\cal M}_3 $ corresponds to $\epsilon_3^{LLz}$ and $\epsilon_3^{RRz}$
($J_{V\mp A}J_{V\mp A}$), the positive sign to
$\epsilon_3^{LRz}$ and $\epsilon_3^{RLz}$
($J_{V\mp A}\-J_{V\pm A}$).
The sign of ${\cal M}_4 $ is positive for
the combinations $\epsilon_4^{LLL}$, $\epsilon_4^{RRL}$,
$\epsilon_4^{RLR}$, $\epsilon_4^{LRR}$ ($J_{V \mp A}J_{TL/TR}j_{V-A}$
and $J_{V \pm A}J_{TL/TR}j_{V+A}$).
For the combinations $\epsilon_4^{LLR}$, $\epsilon_4^{RRR}$,
$\epsilon_4^{RLL}$, $\epsilon_4^{LRL}$
($J_{V \mp A}J_{TL/TR}j_{V+A}$
and $J_{V \pm A}J_{TL/TR}j_{V-A}$) it is negative.
The sign of the matrix element ${\cal M}_5 $ is negative for the left-handed
leptonic current ($\epsilon_5^{xyL}$) and positive for the right-handed one
($\epsilon_5^{xyR}$).

The standard nuclear matrix elements
${\cal M}_{F,N}$ und ${\cal M}_{GT,N}$
in eq. \rf{nucl1} are defined as

\be{mgth}
{\cal M}_{GT,N} =
\big< F | \sum_{i \ne j} \tau_{+}^{(i)} \tau_{+}^{(j)}
                     \vec{\sigma}_i \cdot \vec{\sigma}_j
                     \left(\frac{R_0}{r_{ij}}\right)
                      F_{N}(x_{A})                | I \big>,
\ee
\be{mfh}
{\cal M}_{F,N} =
                 \big< F | \sum_{i \ne j} \tau_{+}^{(i)} \tau_{+}^{(j)}
                     \left(\frac{R_0}{r_{ij}}\right)
                      F_{N}(x_{A})                | I \big>.
\ee
Here $r_{ij}$ and  $R_0$ are the distance of the nucleons involved and the
nuclear radius respectively.
Numerical values of these matrix elements for $^{76}$Ge \cite{Hir95b}
in the Quasi Particle Random Phase Approximation(pn-QRPA) are given in
Table 1. Uncertainties of nuclear matrix element calculations are discussed
in Ref. \cite{Hir95b,mink}.

The prefactors $\alpha_i$ in eq. \rf{nucl1}
are defined as follows,
\ba{alpha}
\alpha_1=\Big(\frac{F_S^{(3)}}{g_A}\Big)^2 \cdot\frac{m_A^2}{m_P m_e},\ \ \
\alpha_2=8 \Big(\frac{ T_1^{(3)}}{g_A}\Big)^2 \cdot\frac{m_A^2}{m_P m_e},\\
\nn
\alpha_3=\Big(\frac{g_V}{g_A}\Big)^2,\ \ \
\alpha_4=\frac{T_1^{(3)}}{g_A} \cdot\frac{m_A^2}{m_P m_e}\ \ \
\alpha_5=\frac{g_V F_S^{(3)}}{g_A^2} \cdot\frac{m_A^2}{m_P m_e}.
\ea
The finite nucleon size is taken into account in a common way \cite{verg} by
introducing the nucleon form factors in a dipole form
\be{formeq}
\frac{g_{V,A}(q^2)}{g_{V,A}}=\frac{F_S(q^2)}{F_S}=
\frac{T_1^{(3)}(q^2)}{T_1^{(3)}}=\left(1-\frac{q^2}{m_A^2}\right)^{-2},
\ee
with $m_A=0.85$ GeV, $g_V= 1.0, g_A= 1.26$.
The other form factor normalizations have been calculated in Ref. \cite{adl}
within the MIT bag model,
\ba{num2}
F_S^{(3)}= 0.48,\ \ \
T_1^{(3)}= 1.38.
\ea
The momentum dependence has been absorbed into the integral
over the momentum $q$ transferred between two nucleons,
\be{d1}
F_N (x_A) = 4 \pi m_{A}^{6}  {r}_{ij} \int \frac{d^3 \vec{q}}{(2\pi)^3}
\frac{1}{(m_A^2 + \vec{q}^2)^4}
        e^{i \vec{q} \vec{r}_{ij}} = \frac{x_A}{48}(3+3x_A+x_A^2)e^{-x_A},
\ee
where $x_A=m_A r_{ij}$.

\section{General \znbb-decay constraints}
Let us extract limits on the lepton-number violating
parameters $\epsilon_k$ in eq. \rf{t12} from the experimental lower bound
on the \znbb-decay half-life. The conservative half-life limit of
the Heidelberg--Moscow experiment
\cite{HM97}
\ba{exp}
T_{1/2}^{0\nu\beta\beta}>1.8 \cdot 10^{25} years (90 \% C.L.)
\ea
provides at present the most stringent bound.
Applying this limit to eq. \rf{t12} leads to a complex
5-dimensional exclusion plot for all five parameters $\epsilon_k$.
To make the \znbb-decay
experimental bound more transparent it
is a common practice to assume the dominance of one coupling.
This means only one non-zero $\epsilon_i$ is considered at  a time
(evaluation "on axis"), while the interference
between different contributions is neglected.
In this way we obtain the upper bounds on $\epsilon_k$
listed in Table 2.

In the general case, when more than one
$\epsilon_{\alpha}$ is involved simultaneously,
we deal with the multidimensional exclusion plot.
For example, this situation occurs in the left-right
symmetric models, see \cite{Hir96c,Doi85}. It also appears natural, if
effective couplings are generated in Fierz rearrangements or decompositions
of given expressions. In the following, as an example, we dicuss the case
of interference
of $\epsilon_4^{LRR}$ and $\epsilon_5^{LRR}$
contributions (the cut in the
$\epsilon_4^{LRR}$-$\epsilon_5^{LRR}$-plane), which can result from the
decomposition
$\bar{u} \gamma_\mu \gamma_\nu (1+\gamma_5)d J^{\mu} j^{\nu}=
(g_{\mu\nu} J  - i J_{\mu\nu}) J^{\mu} j^{\nu}$ mentioned above.
In this case the expression for the inverse half life becomes
\be{}
[T_{1/2}^{0\nu\beta\beta}]^{-1}=|\epsilon_4^{LRR} {\cal M}_4
+ \epsilon_5^{LRR} {\cal M}_5|^2
G_2,
\ee
leading to the exclusion plot shown in Fig. 2. The region outside the
ellipse is excluded. It is obvious that the limits on $\epsilon_4^{LRR}$ and
$\epsilon_5^{LRR}$ in this case are somewhat less stringent (by about 15-20\%)
than the on-axis limits listed in Table 2. The analogous treatment of other
combinations, using the matrix elements listed in Table 1, is
straightforward.

\section{Conclusion}
We have presented a general parameterization for the short range part
of the
neutrinoless double beta decay rate in terms of effective couplings.
The resulting bounds are summarized in Table 2.
Combined with the long range part already published \cite{supf},
this parameterization provides
the double beta decay constraints
for arbitrary lepton number violating theories beyond the SM.

\section*{Acknowledgement}
M.H. would like to acknowledge support by the European Union's
TMR program under grant ERBFMBICT983000. H.P. wants to thank the members of
the Instituto de Fisica Corpuscular (C.S.I.C.) Valencia for their kind
hospitality. S.G.K. was supported in part by Fondecyt (Chile) under
grant 1000717.

\newpage
\nopagebreak

\begin{figure}
\epsfysize=45mm
\epsfbox{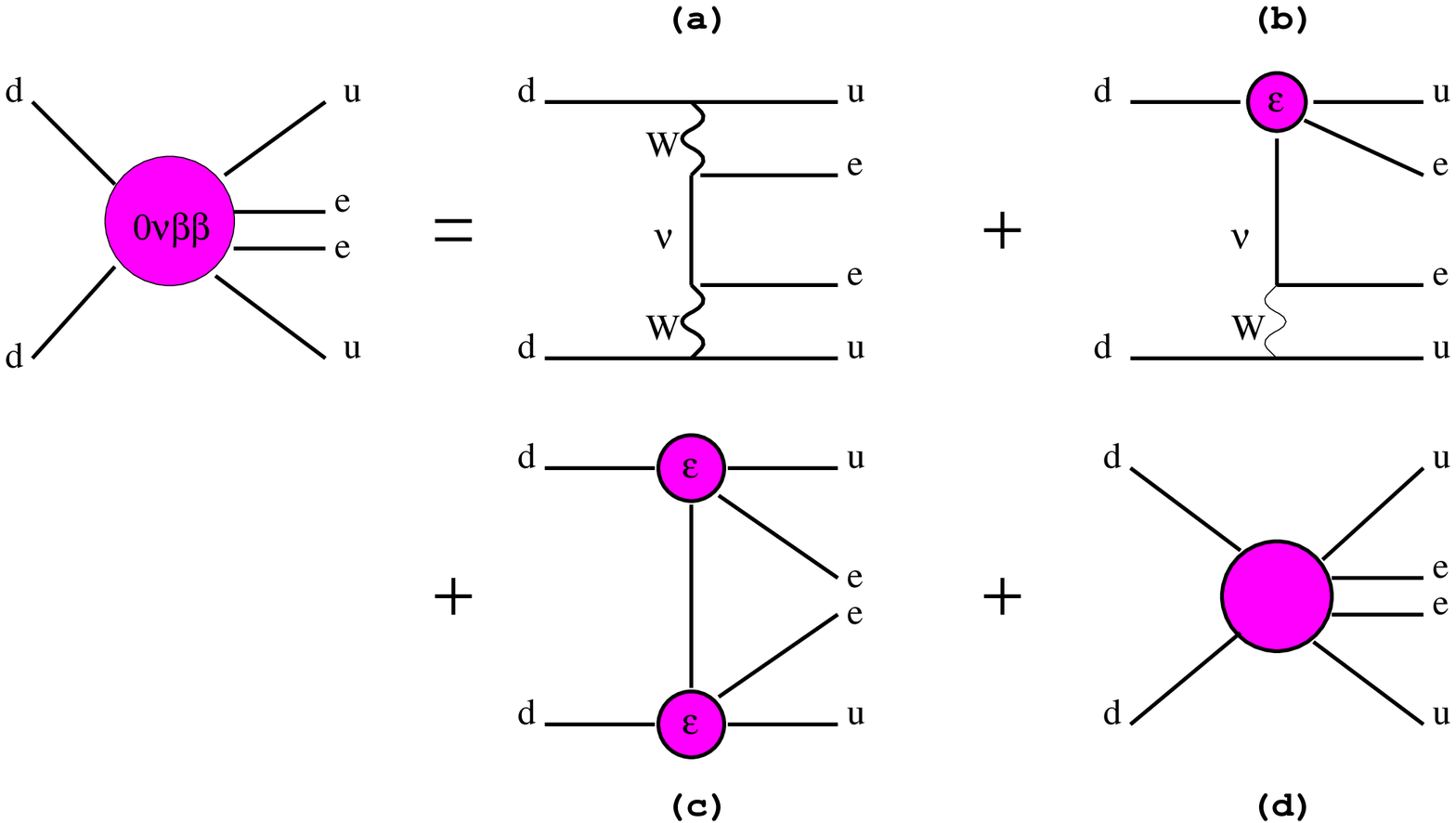}
\caption{\it Feynman graphs of the general double beta rate:
The contributions a) - c)
correspond to the long range part discussed in \protect{\cite{supf}},
the contribution d) is the short range
part.}
\label{1}
\end{figure}

\begin{figure}
\epsfysize=100mm
\epsfbox{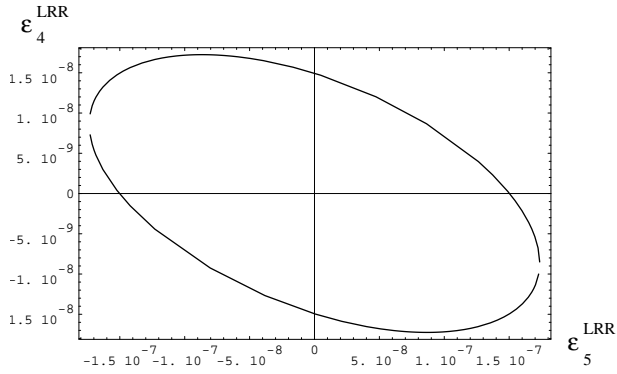}
\vspace*{-20mm}
\caption{\it
Simultaneous treatment of $\epsilon_4^{LRR}$ and
$\epsilon_5^{LRR}$. The $0\nu\beta\beta$ half life
limit allows for a region (inside the ellipse)
in a two dimensional parameter space.
The limits on
$\epsilon_4^{LRR}$ and $\epsilon_5^{LRR}$ in this case are somewhat
(by about 15-20\%)
less stringent than the ones evaluated ``on axis''.
}
\label{1}
\end{figure}

\vspace*{2cm}

\newpage

\begin{table}

\bigskip

\begin{tabular}{c|c|c|c|c}
\hline
\hline
${\cal M}_{GT,N}$
&
${\cal M}_{F,N}$
&
$G_{01}(yr)^{-1}$
&
$G_{06}(yr)^{-1}$
&
$G_{09}(yr)^{-1}$ \\
\hline
$1.13 \cdot 10^{-1}$ &
$4.07 \cdot 10^{-2}$ &
$6.4 \cdot 10^{-15}$&
$1.43\cdot 10^{-12}$&
$3.3 \cdot 10^{-10}$ \\
\hline
\hline
\end{tabular}
\vspace*{3mm}
\caption{\it  Nuclear matrix elements for  $^{76}$Ge \znbb-decay
calculated in the pn-QRPA approach
(from \protect\cite{Hir95b}) and phase space
factors (from \protect\cite{Doi85}). }
\end{table}

\begin{table}

\bigskip

\begin{tabular}{c|c|c|c|c|c}
\hline
\hline
$|\epsilon_1|$ &
$|\epsilon_2|$ &
$|\epsilon_3^{LLz}|,|\epsilon_3^{RRz}|$   &
$|\epsilon_3^{LRz}|,|\epsilon_3^{RLz}|$   &
$|\epsilon_4|$ &
$|\epsilon_5|$ \\
\hline
 $3 \cdot 10^{-7}$ &  $2 \cdot 10^{-9}$ &  $4 \cdot 10^{-8}$
&  $1 \cdot 10^{-8}$ & $2 \cdot 10^{-8}$ & $2 \cdot 10^{-7}$ \\
\hline
\hline
\end{tabular}
\vspace*{3mm}
\caption{\it Experimental upper bounds on the absolute
values of the effective $B-L$ violating couplings evaluated ``on axis''.
For $\epsilon_3$ the contractions of hadronic currents with
different chiralities lead to different results.}
\end{table}
\end{document}